\documentclass[prd,twocolumn,floats,floatfix,showpacs,nofootinbib]{revtex4}
\usepackage{graphicx}
\usepackage{dcolumn}
\usepackage{bm}
\usepackage{graphics}
\usepackage{slashed}
\usepackage{amssymb}
\usepackage{natbib}
\usepackage{amsmath}
\usepackage{amsthm}
\usepackage{url}
\usepackage{color}
\newcommand{\fracc}[2]{\frac{\textstyle{#1}}{\textstyle{#2}}}

\begin{document}

\title{The gravitational energy}

\author{M. Novello}\email{novello@cbpf.br}
\author{E. Bittencourt}\email{eduhsb@cbpf.br}
\affiliation{Instituto de Cosmologia Relatividade Astrofisica ICRA -
CBPF\\ Rua Dr. Xavier Sigaud 150 - 22290-180 Rio de Janeiro - Brazil\\}


\date{\today}

\begin{abstract}
We present the expression $ t_{\mu\nu} $ of the energy-momentum
tensor of the gravitational field in the framework of the recent
proposal of the Geometric Scalar theory of gravity (GSG). From the
conservation of $t_{\mu\nu} $ it follows the dynamics of the
gravitational field. As an example of this expression for
$t_{\mu\nu} $  we calculate the gravitational energy of a compact
object.
\end{abstract}

\maketitle

\section{Introduction}

In spite of several attempts, the theory of general relativity (GR)
is not able to exhibit a true energy-momentum tensor for the
gravitational field. This is usually understood as the impossibility
to localize the gravitational energy. To bypass this criticism, many
proposals have been discussed in the literature suggesting that for
the case of gravity the ambiguity of the localization of its energy
should be necessarily treated in terms of pseudo-tensorial
quantities as it was proposed by Einstein \cite{einstein}, Tolman
\cite{tolman}, Landau and Lifshitz \cite{landau} and M{\o}ller
\cite{moller}\footnote{ See the analysis in \cite{paulo} and
references therein. An alternative point of view was proposed by Bel
 \cite{bel}.}. The origin of such difficulty is due to the characteristic of
this theory that deals with a dynamics imposed directly on the
geometry. Such property led to the general acceptance that for
gravity processes, the definition of energy should be transcended
and considered as a fruitful and important concept limited to some
special configurations like, for instance, in asymptotically flat
geometries.

This situation is not present in those alternative theories in
which the dynamics of the gravitational field is not directly
imposed upon the metric structure. This is the case of the recent
Geometric Scalar Gravity (GSG) \cite{novellojcap13} that combine the main idea of general
relativity---that is, gravity is a geometrical phenomenon that
should be treated as a modification of the metric of
space-time---with the dynamics being imposed only indirectly to the
geometry. The construction of a true energy-momentum tensor for the
gravitational field in the GSG is the focus of this paper.

We shall present arguments to support the
identification of a symmetric second order tensor to the
energy-momentum tensor of the gravitational field. This will be made
in the realm of the GSG
in which the basic fundamental quantity that represents the
gravitational field is the scalar field $ \Phi.$

The characterization of the gravity as nothing but the geometry of
space-time gave a beautiful interpretation of the equivalence
principle. The fact that it is possible to annihilate locally the
gravitational effects led to the general belief that the
gravitational energy should not be associated to a true tensorial
object but to a pseudo-tensor in order to be eliminated locally
according to the equivalence principle.

The discovery of a pseudo-tensor to represent the non-localization
of the gravitational energy was perfectly in accord with these main
ideas that are at the basis of general relativity. Einstein for
instance proposed to characterize the energy content in the
gravitational field in terms of a super-potential $
H_{\mu}{}^{[\nu\lambda]} $ such that one obtains the pseudo-tensor
$$T^{\nu}{}_{\mu} = \partial_{\lambda} H_{\mu}{}^{[\nu\lambda]},$$
where
$$ H_{\mu}{}^{[\nu\lambda]}  = \frac{1}{2 \kappa} \, \widetilde{g}_{\mu\varepsilon} \, \partial_{\alpha}
\left(\widetilde{g}^{\varepsilon\nu} \,
\widetilde{g}^{\alpha\lambda} - \widetilde{g}^{\varepsilon\lambda}
\, \widetilde{g}^{\alpha\nu} \right).$$

We note in passing that distinct from all other energy-momentum
tensor of field theories---e.g., electromagnetic field---that
contains only first order derivative of the field, the role of the
energy-momentum tensor displayed in GR by the quantity $
T^{\nu}{}_{\mu}$ contains second order derivatives of the basic
variable $\widetilde{g}^{\alpha\lambda} = \sqrt{- g} \,
g^{\alpha\lambda}.$

Nothing similar in the Geometric Scalar Gravity that attributes a
real tensor $t^{\mu\nu}_{g} $ to the description of the energy
content of the field and consequently cannot be locally made to
vanish by any choice of coordinates. We emphasize that this property
does not conflict with the Equivalence Principle, once the
basic variable $ \Phi$ interacts with matter only through the
combination that generate the metric $ q_{\mu\nu}.$ The universality
of the gravitational interaction of any kind of matter and energy
only through this metric and its derivatives allows the possibility
of local annihilation of the gravitational field by making to vanish
the associated Christoffel symbol. This makes GSG in perfect
agreement with the Equivalence Principle.

The adjective ``geometric" in its name means that this theory accepts the
hypothesis originally made in general relativity, that gravity is a
geometrical phenomenon described by a Riemannian metric. In GR
the ten components of the metric tensor are the basic variables of
the theory (up to coordinate transformations). In GSG the metric
tensor is determined by the derivatives of a fundamental independent
physical quantity represented by the scalar field $ \Phi.$ In order
to exhibit the similarities and differences between GSG and GR let us
summarize their main properties.\\

{\bf Basic properties of General Relativity:}

\begin{itemize}
\item{The gravitational interaction is described by a
second order tensor field $ h_{\mu\nu} $;}
\item{The field $h_{\mu\nu}$ satisfies a nonlinear dynamics;}
\item{The theory satisfies the principle of general covariance which,
in other words, means that GR is not a theory restricted to the realm of
special relativity;}
\item{All kinds of matter and energy interact with $h_{\mu\nu}$
only through the pseudo-Riemannian metric
$$g^{\mu\nu} =  \eta^{\mu\nu} + h^{\mu\nu};$$}
\item{Test particles follow geodesics relative to the gravitational metric $ g_{\mu\nu};$}
\item{ $h_{\mu\nu}$ is related in a nontrivial way with the Newtonian potential $\Phi_N$;}
\item{The background Minkowski metric
is not observable. Matter and energy interact gravitationally only
through the combination $ \eta_{\mu\nu}+ h_{\mu\nu}$ and its
derivatives;}
\item{Electromagnetic waves propagate along null geodesics relative to the metric $ g_{\mu\nu}.$}
\item{The contravariant definition of the metric in GR, that is,
$$g^{\mu\nu} =  \eta^{\mu\nu} + h^{\mu\nu}$$ is an exact expression.
Thus the corresponding co-variant expression, defined as the inverse
$g_{\mu\nu} \, g^{\nu\lambda} = \delta^{\lambda}_{\mu}$, is given by
the infinite series
$$ g_{\mu\nu} = \eta_{\mu\nu} - h_{\mu\nu} + h_{\mu\alpha} \,
h^{\alpha}{}_{\nu} + ... $$}
\end{itemize}

{\bf Basic properties of the Geometric Scalar Gravity:}

\begin{itemize}
\item{The gravitational interaction is described by a
scalar field $ \Phi$;}
\item{The field $\Phi$ satisfies a nonlinear dynamics;}
\item{The theory satisfies the principle of general covariance which,
in other words, means that GSG is not a theory restricted to the realm of
special relativity;}
\item{All kinds of matter and energy interact with $\Phi$
only through the pseudo-Riemannian metric
\begin{equation}
q^{\mu\nu} = \alpha \, \eta^{\mu\nu} + \frac{\beta}{w} \,
\partial^{\mu}\Phi \,\partial^{\nu} \Phi,
\label{4jul}
\end{equation}
where $ \alpha$ and $\beta$ are functions of $\Phi$ and $w \equiv
\partial_{\mu}\Phi \,\partial_{\nu} \Phi \, \eta^{\mu\nu};$}
\item{Test particles follow geodesics relative to the gravitational metric $ q_{\mu\nu};$}
\item{ $\Phi$ is related in a nontrivial way with the Newtonian potential $\Phi_N$;}
\item{The background Minkowski metric
is not observable. Matter and energy interact gravitationally only
through the combination $$ \alpha\, \eta^{\mu\nu} + \frac{\beta}{w}
\,
\partial^{\mu}\Phi \,\partial^{\nu} \Phi$$ and its derivatives;}
\item{Electromagnetic waves propagate along null geodesics relative to the metric $ q^{\mu\nu};$}
\item{The contravariant definition of the metric in GSG, that is,
equation (\ref{4jul}) is an exact expression. The corresponding
covariant expression, defined as the inverse $q_{\mu\nu} \,
q^{\nu\lambda} = \delta^{\lambda}_{\mu}$, is also a binomial
expression
\begin{equation}
q_{\mu\nu} = \frac{1}{\alpha} \, \eta_{\mu\nu} - \frac{\beta}{\alpha
\, (\alpha + \beta) \, w} \, \partial_{\mu} \Phi \, \partial_{\nu}
\Phi. \label{291} \end{equation} }
\end{itemize}

\vspace{0.50cm}

The  parameters $ \alpha $ and $ \beta$ that are functionals of the
scalar field $ \Phi$ were specified in \cite{novellojcap13} by
fixing the dynamics of the scalar field. We note that in both
theories (GR and GSG) the auxiliary (Minkowski) metric $
\eta^{\mu\nu}$ is unobservable because the gravitational field
couples to matter only through the effective metric $ g^{\mu\nu}$ in
the case of GR, or to $ q^{\mu\nu}$ in the case of GSG. Thus, in
both theories a unique geometrical entity interacts with all forms
of matter and energy and the geometry underlying all events is
controlled by the gravitational phenomena.

From this postulate it follows immediately that the space-time
geometry is an evolutionary process identified to the dynamics of
the gravitational field. This beautiful hypothesis made by Einstein
and that constitutes the true basis of general relativity is
contained in each observation as a specific example of a geometry
solving that dynamics. That is, the metric couples universally and
minimally to all fields of the standard model by replacing
everywhere the Minkowski metric $ \eta_{\mu\nu}$ either by  $
g_{\mu\nu},$ in the case of GR, or by  $ q_{\mu\nu}$ in the case of
GSG.

The origin of this departure from general relativity was explained in
\cite{novellojcap13} where the expressions of $\alpha$ and
$\beta$ were obtained from the analysis of the gravitational
properties of planetary orbits yielding

\begin{equation}
\alpha = e^{- 2\, \Phi}, \label{13jul1}
\end{equation}
\begin{equation}
\beta = \frac{1}{4}( e^{-2\Phi} - 1)(e^{-2\Phi}-9) \label{13jul2}
\end{equation}
and
\begin{equation}
V = \frac{(e^{\Phi} - 3e^{3\Phi})^{2}}{4}, \label{13jul222}
\end{equation}
where $V (\Phi) $ is the nonlinear potential of the lagrangian that
is at the basis of GSG.

We shall prove that in the Geometric Scalar Gravity, the energy density of a
star---let us call it $ \mathcal{E}_{g}$---is obtained as a solution
of the first-order differential equation

$$ \frac{d \mathcal{E}_{g}}{d \Phi} + Q(\Phi) \, \mathcal{E}_{g} + P(\Phi) = 0, $$
where the functions $ Q $  and $ P $ are given in terms of the
scalar gravitational field. The gravitational energy density
$\mathcal{E}_{g}$ is defined as the $0-0$ component of the true
energy-momentum tensor $ t^{\mu\nu}_{g}.$ The simplest way to
present this proposal is to use the description of GSG in the
framework of a field theory suggested by Gupta \cite{gupta}, Deser \cite{deser},
Grishchuk \cite{grish}, Feynman \cite{feynman} and
others in the case of GR. We start this program by a short review of
the correspondent scheme of general relativity.

\section{The field theory description of general relativity}

In the geometric formulation of general relativity the metric of
space-time $ g_{\mu\nu}$ satisfies the equation
\begin{equation}
G_{\mu\nu} \equiv R_{\mu\nu} - \frac{1}{2} \, R \, g_{\mu\nu} = -
\kappa \, T_{\mu\nu}.
\label{GR}
\end{equation}

There is another form to present such dynamics using a description
in terms of a field theory embedded in an auxiliary flat space-time
taken a priori as non-observable \cite{feynman}.

In this formulation, one starts by defining a symmetric second order
tensor $ h_{\mu\nu}$, which lies on the Minkowski background $
\eta_{\mu\nu}.$ Then the hypothesis is made that matter and energy
of any kind interacts gravitationally only through the
combination\footnote{Let us note that Grishchuk {\it et al.} \cite{grish} made a
different choice and they use the definition not in terms of a
tensorial equation, but defining the pseudo-tensor terms as $$
\sqrt{- g} \, g^{\mu\nu} = \sqrt{- \eta} \, ( \eta^{\mu\nu} +
h^{\mu\nu}).$$} given by equation

\begin{equation}
g^{\mu\nu} = \eta^{\mu\nu} + h^{\mu\nu}. \label{220833}
\end{equation}

This binomial form is an exact expression for the metric $
g^{\mu\nu}.$ Consequently the covariant tensor $g_{\mu\nu}$ is an
infinite series:
\begin{equation}
g_{\mu\nu} = \eta_{\mu\nu} -
h_{\mu\nu} + h_{\mu\alpha} \, h^{\alpha}{}_{\nu} + ...
\label{220831}
\end{equation}

There are two main postulates founding general relativity:
\begin{itemize}
\item{The background Minkowski metric
is not observable. Matter and energy interact gravitationally only
through the combination $ \eta^{\mu\nu}+ h^{\mu\nu}$ and its
derivatives. Any test body in a gravitational field moves along a
geodesic relative to the metric $g_{\mu\nu}$;}
\item{The dynamics of gravity is described by an equation relating
the contracted curvature tensor $ R_{\mu\nu} $ to the stress-energy
tensor of matter.}
\end{itemize}

In the next section, these postulates will be applied in the
derivation of the field theory formulation of GR.

\section{Interpreting the non-linear terms of the dynamics of the
gravitational field $ g_{\mu\nu} $ of General Relativity in terms of
equivalent ``energy-momentum representation"}

Using equations (\ref{220833}) and (\ref{220831}) we re-write the
dynamics (\ref{GR}) as an infinite series

\begin{equation}
G_{\mu\nu}^{L} = - \kappa \, \left( T_{\mu\nu} + t_{\mu\nu}^{(1)} +
t_{\mu\nu}^{(2)} + t_{\mu\nu}^{(3)} + ...\right), \label{21086}
\end{equation}
where $G_{\mu\nu}^{L}$ is the linear part of the Einstein tensor $
G_{\mu\nu} $ and the quantities $ t_{\mu\nu}^{(n)}$  for $n = 1, 2,
3, ...$ contains non-linear terms of the Ricci tensor developed in
series of order $ h^{n}.$ The tensor of matter is divergence free
$$ T^{\mu\nu}{}_{;\nu} = 0.$$

Writing in a compact form
$$ G_{\mu\nu}^{L} = - \, \kappa \, \left( T_{\mu\nu} + t_{\mu\nu}
\right), $$ we note that although the quantity $t_{\mu\nu}$ appears
as the non-linear source of the gravitational field $ h_{\mu\nu} $
it should not be identified with the gravitational energy, once it
follows from this dynamics that it is not a conserved tensor in the
metric $ g_{\mu\nu}, $ once its divergence does not vanish
$$ t^{\mu\nu}{}_{;\nu} \neq 0,$$
which is the origin of the main difficulties to undertake this
field-theoretical path to construct a well-grounded energy-momentum
tensor for the gravitational field within the GR theory.

\section{The field theory description of the geometrical scalar gravity}

The dynamics of the GSG as proposed in \cite{novellojcap13} is given
by the equation
\begin{equation}
\sqrt{V} \, \Box\Phi=  \kappa \, \chi, \label{12out1}
\end{equation}
where $ V $ is given by equation (\ref{13jul222}) and the d'Alembert
operator $ \Box $ is defined in the curved metric
$$ q^{\mu\nu} = \alpha \, \eta^{\mu\nu} + \beta \, \frac{\partial^{\mu} \Phi \,
\partial^{\nu} \Phi}{w},$$ that is
\begin{equation}
\Box \Phi \equiv \frac{1}{\sqrt{- \, q}} \, \partial_{\mu} \left( \sqrt{- \, q}
\,\partial_{\nu} \Phi \, q^{\mu\nu}\right) . \label{1093}
\end{equation}
The determinant $q$ of $q_{\mu\nu}$ is given by
\begin{equation}
\sqrt{- \, q} = \frac{\sqrt{- \, \eta}}{ \alpha^{3/2} \,
\sqrt{\alpha + \beta}}. \label{4091} \end{equation} The source $\chi
$ of the gravitational field was obtained in \cite{novellojcap13}
and its explicit expression is written as
\begin{equation} \chi = \fracc{1}{2} \, \left( \frac{3 \, e^{2
\Phi} + 1}{3 \, e^{2\Phi}- 1} \, E - T - \nabla_{\lambda} \,
C^{\lambda}\right). \label{20081} \end{equation} The quantities $ E$, $T$
and $C^{\lambda} $ are given in terms of the energy-momentum tensor
$$ T\equiv T^{\mu\nu} \, q_{\mu\nu}, \hspace{.3cm}
E \equiv \frac{T^{\mu\nu} \, \partial_{\mu}\Phi \,
\partial_{\nu}\Phi}{\Omega} $$ and
$$ C^{\lambda}\equiv\frac{\beta}{\alpha \, \Omega} \, \left( E \, q^{\lambda\mu} - T^{\lambda\mu} \right) \, \partial_{\mu}\Phi, $$
where $$ \Omega \equiv \partial_{\mu} \Phi \partial_{\nu}\Phi \,
q^{\mu\nu} = \alpha^{3} \, V \, w.$$ The energy-momentum tensor of
matter is defined in the standard way
\begin{equation}
T_{\mu\nu} = \frac{2}{\sqrt{-q}} \, \frac{\delta (\sqrt{-q} \,
L_{mat})}{\delta q^{\mu\nu}}. \label{22084}
\end{equation}

Equation (\ref{12out1}) describes the dynamics of the GSG in
presence of matter. The source of the gravity expressed by the
quantity $\chi$ involves a non-trivial coupling between the gradient
of the scalar field $\partial_{\mu} \Phi$ and the energy-momentum
tensor of the matter field $T_{\mu\nu}$ and not uniquely its trace,
as the previous unsuccessful scalar theories of gravity \cite{2}.

In order to express this theory along the same lines as it was made
in the case of general relativity, let us describe the dynamics of
GSG following similar steps, according to Feynman's approach, and
try to describe the dynamics of the scalar field $\Phi$ in a flat
space-time endowed with the metric $ \eta_{\mu\nu}.$  We start with
 the Lagrangian
\begin{equation}
L = V(\Phi) \, w.
\label{1091}
\end{equation}
Variation of $ L $ with respect to $ \Phi$ yields
$$ \delta \,  \int \, \sqrt{- \eta} \, L = \int \, \sqrt{- \eta} \,
\left( V \, \Box_{M} \Phi + \frac{1}{2} \, \frac{dV}{d\Phi} \,w
\right) \, \delta \Phi. $$ In this expression $ \Box_{M} $
represents the d'Alembert operator in the auxiliary Minkowski
metric, defined as
$$\Box_{M} \Phi = \frac{1}{\sqrt{- \eta}} \, \partial_{\mu} \left( \sqrt{-
\eta} \,\eta^{\mu\nu}\,\partial_{\nu} \Phi\right). $$

The dynamics issued from this Lagrangian is given by
\begin{equation}
V \,\Box_{M} \Phi + \frac{1}{2} \, V'\, w = 0,
\label{19081}
\end{equation}
where $V'\equiv dV/d\Phi$. Using the identity
\begin{equation}
\label{brid_rel}
\Box \Phi = (\alpha +\beta) \,\left(\Box_{M} \Phi + \frac{1}{2} \,
\frac{V'}{V} \, w\right),
\end{equation}
where $ \Box$ is given by (\ref{1093}),
it follows that the dynamics described by (\ref{1091}) is equivalent
to the form
\begin{equation}
\Box \Phi =0. \label{19082}
\end{equation}

At this point, the geometrical scalar gravity takes a step beyond
and following similar lines as in GR states the fundamental
hypothesis that all forms of matter and energy interact with the
gravitational field $ \Phi $ {\bf{only}} through the combination of
metric $ q_{\mu\nu}$ and its derivatives in a covariant way. In
other words, given the Lagrangian of matter $L_{mat}$ in the special
relativistic theory then, in order to couple this matter to the
gravitational field, one has only to use the minimal coupling
principle and substitute the unobserved metric $ \eta_{\mu\nu} $ by
the gravitational one $ q_{\mu\nu}.$ Thus, using this principle one
obtains the equation of motion that drives the effects of matter in
the generation of a gravitational field, that is equation
(\ref{12out1}). The final step to complete the theory is to fix the
dependence of $ V $ on the coefficients $ \alpha$ and $\beta$ and
its relation with $ \Phi.$ In the quoted paper we obtained the
values displayed above in Eqs.\ (\ref{13jul1}), (\ref{13jul2}) and
(\ref{13jul222}).

Then, inserting these expressions in the matter
action $\int \sqrt{- q} \, L_{mat}$ one obtains the form of $
\chi$ given above in Eq.\ (\ref{12out1}).

Let us remark that only the trace $ T $ and the projections of $
T_{\mu\nu}$ onto the gradient of $ \Phi $ appear in this expression.
This means that only five components from the ten contained in the
energy-momentum tensor of matter appear in the dynamics. Nothing
similar in GR once in this theory all components of $ T_{\mu\nu}$
are involved in the dynamics.

\section{Interpreting the non-linear terms of the dynamics of the
gravitational field $ \Phi $ in terms of equivalent ``energy-momentum
representation"}

In this section we will finish the task of achieving in the
Geometric Scalar Gravity the equivalent result of Gupta-Feynman
representation of GR as a field theory description in an
unobservable background endowed with a Minkowski metric.

From the bridge relation (\ref{brid_rel}), we rewrite
the dynamics of $ \Phi $ in terms of the auxiliary flat space-time.
The dynamics of $ \Phi$ in (\ref{12out1}) takes the form
\begin{equation}
\Box_{M} \Phi = \frac{\kappa \, \chi_{mat}}{(\alpha + \beta) \,
\sqrt{V}} \, - \frac{V'}{2 \,V} \, w
\end{equation}
where $ \chi_{mat}$ represents the source matter terms. Our task
then is to construct an associated energy-momentum tensor $
\Theta_{\mu\nu}$ such that using the expression (\ref{20081}), we
can re-write the non-linear term of the r.h.s.\ of the above
equation as a source of the field or, in other words, to choose $
\Theta_{\mu\nu} $ such that the $ \chi-$term associated to this
tensor reproduces precisely the non-linear term that is $$
\frac{\kappa \, \chi_{\theta}}{(\alpha + \beta) \, \sqrt{V}}  = - \,
\frac{V'}{2V} \, w.$$

A first guess is almost univocally determined by setting
\begin{equation}
\Theta_{\mu\nu}
= a \, \partial_{\mu} \Phi \, \partial_{\nu} \Phi + b \,
q_{\mu\nu},
\label{20082}
\end{equation}
where $ a $ and $ b $ may depend on $ \Phi$ and on its derivative $ \Omega.$
Let us prove that this is indeed possible. From the definitions of
$ E, T$ and $ C^{\lambda}$ a direct calculation yields
$$ E_{\Theta} = a \, \Omega + b ;\quad T_{\Theta} = a \, \Omega + 4 \, b \quad \mbox{and}\quad (C^{\lambda})_{\Theta} = 0.$$
Using these values on the expression of $ \chi$ we obtain
\begin{equation}
\chi_{\Theta} = \frac{1}{2 \, (3 - \alpha)} \, [ 2 \, \alpha
\,\Omega \,a + ( 5 \, \alpha -9) \,b]. \label{20083}
\end{equation}
The question now is to find values of $ a $ and $ b $ such that
allows for the identification:
$$ - \, \frac{w \, V'}{2 \, V} = \kappa \, \frac{\chi_{\Theta}}{(\alpha +
\beta) \, \sqrt{V}}.$$ Developing both sides of this equation we
obtain the unique condition relating $ a $ and $ b $, that is:
\begin{equation}
b = \frac{2 \, \Omega}{9 \, e^{2 \Phi} - 5} \, \left( a - \sqrt{V}
\, (1 - 9 \, e^{2 \Phi}) \right) \label{21081}
\end{equation}
A particular solution of this relation is given by the values
$$ a = - 4 \, \sqrt{V} \quad \mbox{and} \quad b = 2 \, \sqrt{V} \, \Omega.
$$ (In the appendix [\ref{app1}] we explain the origin of such choice.)

This ends the proof that it is
possible to describe the dynamics of GSG through a similar procedure
as it was made by GR, that is we can rewrite the gravitational
dynamics (\ref{12out1}) under the form
\begin{equation}
\Box_{M} \Phi = \frac{\chi_{mat} + \chi_{\theta}}{(\alpha + \beta)
\, \sqrt{V}}, \label{21084}
\end{equation}
which is the version for GSG of the expression (\ref{21086}) of
general relativity.

The first part of our task is finished. Let us now turn to a more
ambitious question, that is, such symmetric second order tensor $
\Theta_{\mu\nu} $ can be associated to the gravitational energy?

In the absence of matter we should expect that the true
energy-momentum tensor of the gravitational field $ t^{\mu\nu}_{g} $
 should be conserved, that is, we should have
$$ t^{\mu\nu}_{g}{}_{; \nu} = 0 $$ where the covariant derivative, of
course, being taken in the observable gravitational metric $
q_{\mu\nu}. $ In the case of general relativity is evident that the
quantity $ t_{\mu\nu}$ introduced by Feynman and others cannot
represent a conserved energy tensor. What about the quantity $
\Theta_{\mu\nu} $ of GSG? A direct calculation yields

\begin{equation}
\Theta_{\mu}{}^{\nu}{}_{; \nu} = ( \partial_{\nu} a \, \Phi^{\nu} +
a \, \Box \Phi ) \, \Phi_{\mu} + \frac{a}{2} \,
\partial_{\mu} \Omega + \partial_{\mu} b.
\label{22087bis}
\end{equation}
Note that the indexes of $\Theta^{\mu\nu} $ must be lowed and raised with the gravitational metric $q^{\mu\nu}$. Using the particular values of $a$ and $ b$
displayed above, it follows (setting $ \Box \Phi = 0$)

\begin{equation}
\Theta_{\mu}{}^{\nu}{}_{; \nu} = - V' \frac{
\Omega}{\sqrt{V}} \,
\partial_{\mu} \Phi.
\label{22087}
\end{equation}
We have pointed out that not all components of the energy-momentum
tensor enter in the dynamics of $ \Phi.$ This property allows us to
ask the following question: is it possible to add to this
$\Theta_{\mu\nu}$ another term---call it $ \Delta_{\mu\nu} $---such
that it does not change the dynamics of $\Phi$ and generate a
conserved quantity? Let us show that this is indeed
possible\footnote{This is nothing but the freedom to add a total
derivative on the Lagrangian. See the Appendix [\ref{app1}] for more details.}.

\subsection*{The extra tensor $ \Delta_{\mu\nu} $}

From what we have shown in the precedent sections, we realize that
there is a freedom on the expression of the gravitational
 energy-momentum tensor that allows one to envisage the possibility
to add to the tensor $\Theta_{\mu\nu}$ another extra term $
\Delta_{\mu\nu} $ that satisfies the two conditions:
\begin{itemize}
\item{It does not affect the dynamics of $ \Phi ;$}
\item{The sum $  \Theta_{\mu\nu} + \Delta_{\mu\nu} $ is
divergence-free.}
\end{itemize}

We set as an attempt, the expression
$$ \Delta_{\mu\nu}  = m \, \partial_{\mu} \Phi \, \partial_{\nu} \Phi + n \,
 q_{\mu\nu} $$ and look for values of $ m $ and $ n $ that satisfies
the above conditions. The associated expression $ \chi_{\Delta} $
defined in terms of the quantities $E, T$ and $C^{\lambda} $ gives
$$ E_{\Delta} = m \, \Omega + n; \quad T_{\Delta} = m \, \Omega +
4 n \quad \mbox{and} \quad C^{\lambda}_{\Delta} = 0. $$

Thus, using equation (\ref{20081}) it follows that
such extra term does not produce any modification on the dynamics
of $ \Phi $ that is, $ \chi_{\Delta}  = 0 $ is

\begin{equation}
\frac{3 \, e^{2 \Phi} + 1}{3 \, e^{2\Phi}- 1} \, E_{\Delta} -
T_{\Delta} = 0, \label{22088}
\end{equation}
which is satisfied by imposing the relation
\begin{equation}
n =\frac{2 \, \Omega}{9 \, e^{2 \Phi} - 5} \, m.
\label{220811}
\end{equation}

Although the tensor $ \Delta_{\mu\nu} $ does not produce any
modification on the dynamics of $ \Phi$, it can help in the
construction of a conserved quantity that we will call the
energy-momentum tensor of the gravitational field:
$$ t^{g}_{\mu\nu} \equiv \Theta_{\mu\nu} + \Delta_{\mu\nu}.$$

Indeed, in order to obtain the dynamics (\ref{19082}) from the
conservation $ t_{g}^{\mu\nu}{}_{ ;  \nu} = 0 $ the condition
\begin{equation}
(a + m)_{, \, \nu} \, \Phi^{\nu} \, \Phi_{\mu} + \frac{1}{2} \, ( a
+ m) \, \Omega_{, \, \mu} + ( b + n)_{, \, \mu} = 0, \label{22089}
\end{equation}
must be satisfied where $a, b, m$ and $n$ are related by the two
equations (\ref{21081}) and (\ref{220811}).

In the particular case of a static and spherically symmetric
configuration admitting a time-like Killing vector where the
gravitational field does not depend on time, the general expression
$$ t^{g}_{\mu\nu} = ( a +m ) \,\Phi_{\mu} \, \Phi_{\nu} + ( b + n ) \,
q_{\mu\nu} $$ provides the total energy of the gravitational field.

We note that what an observer endowed with a four-velocity $v^{\mu}$
call density of energy $\mathcal{E}_{g}$ is given by the projection
$ \mathcal{E}_{g} = t_{\mu\nu} \, v^{\mu} \, v^{\nu}.$ Choose the
normalized four-vector $ v^{\mu} = \sqrt{\alpha} \,
\delta^{\mu}_{0}$ to obtain
\begin{equation}
\mathcal{E}_{g} = b + n,
\label{20911}
\end{equation}
where the density of energy is given by solutions of the equation
(\ref{22089}) that in this case reduces to

\begin{equation}
\label{26083}
\begin{array}{l}
\partial_{\lambda} \,\left( \fracc{(9 \, e^{2\Phi} - 5)}{2 \, \Omega}
\, \mathcal{E}_{g} + \fracc{1}{\kappa} \,\sqrt{V} \,
(1-9 \, e^{2\Phi}) \right) \, \Phi^{\lambda} \, \Phi_{\mu} + \nonumber \\
\fracc{1}{2} \, \left( \fracc{(9 \, e^{2\Phi} - 5)}{2 \, \Omega} \,
\mathcal{E}_{g} +  \fracc{1}{\kappa} \,\sqrt{V} \, (1 - 9 \,
e^{2\Phi}) \right) \, \Omega_{\mu} + \partial_{\mu} \mathcal{E}_{g} = 0.\nonumber\\
\end{array}
\end{equation}

Let us now give a specific example of this formula for the case of a
star.

\section{The gravitational energy of a star}
In \cite{novellojcap13} the gravitational field of a spherically
symmetric and static configuration was calculated. The gravitational
metric associated to this configuration has the same expression as
in the Schwarzschild solution of general relativity. This means that
the geometry constructed with $ q_{\mu\nu} $ in the case that $\Phi
= \Phi(r)$ given by
\begin{equation}
\label{line_el_scha} ds^2 = \left(1-\frac{r_H}{r}\right)dt^2 -
\left(1-\frac{r_H}{r}\right)^{-1}dr^2 - r^2d\Omega^2
\end{equation}
is a solution of both dynamics, that is, the vacuum equations of general relativity
$R_{\mu\nu} = 0$ and also of the geometric scalar theory, that is
\begin{equation}
\Box \Phi =0,
\label{28081}
\end{equation}
where the covariant derivatives are taken, obviously, in the metric
(\ref{line_el_scha}). The gravitational field, satisfying
(\ref{28081}) is
\begin{equation}
\Phi = \frac{1}{2} \, \ln\left(1 - \frac{r_{H}}{r}\right).
\label{220815}
\end{equation}
Let us now evaluate the gravitational energy of a star according to
the analysis presented in the previous section.

The first step is to know the value of tensor $ t_{\mu\nu} =
\Theta_{\mu\nu} + \Delta_{\mu\nu}.$ In the present case a
simplification appears due to the fact that the norm $\Omega$ can be
written as an algebraic expression of the scalar field. Indeed, a
direct calculation gives
$$ \Omega = - \,  \frac{1}{4 \, r_{H}^{2}} \, \frac{(
\alpha - 1 ) ^{4}}{\alpha^{3}}, $$
where $\alpha=(1-r_H/r)^{-1}$. Using this property into equation (\ref{22089}),
we can evaluate the gravitational energy density of a
star by the formula $ \mathcal{E}_{g}= b + n$. From the condition of conservation
(\ref{22089}), we have
$$ \frac{d (a + m)}{d \alpha} \, \Omega + \frac{(a + m )}{2} \,
\frac{d \Omega}{d \alpha} + \frac{d (b + n)}{d \alpha} = 0. $$
Now, from the relation between $ a + m $ and $ b + n $ it follows
$$ a + m = \left( \frac{9 - 5 \alpha}{2 \, \Omega} \right) \,
(b + n) + \frac{\sqrt{V} \,(\alpha - 9)}{\kappa \, \alpha}.$$

Using these results it implies that the gravitational energy density of a
star satisfies the linear differential equation

\begin{equation}
\frac{d \mathcal{E}_{g} }{d \alpha} -  \frac{5 \, \alpha + 3}{6 \,
\alpha \, (\alpha - 1)} \, \mathcal{E}_{g}  + P(\alpha) = 0,
\label{220819}
\end{equation}
where
$$ P(\alpha) = \frac{1}{6 \, \kappa \, r_{h}^{2}} \, \frac{ (\alpha - 1) ^{3} \,(7 \, \alpha^{2} - 45
\, \alpha + 54)}{|\alpha - 3| \, \alpha^{11/2}}.$$ This equation can
be integrated analytically and all details are presented in Appendix
[\ref{app2}]. It is important to remark that we have to choose the
integration constant in order to set $\mathcal{E}_{g}$ equal to zero
when $r$ goes to infinity.

From the conservation of $ t^{\mu\nu}$ it follows that
$$ E_{g} = \int \, t^{\mu 0} \, k_{\mu} \, \sqrt{- q} d^{3}x $$
is a constant, where $ k_{\mu} = (1/\alpha, 0, 0, 0) $ is the
time-like Killing vector of the Schwarzschild solution. Then, we
obtain the gravitational energy $E_g$ for the static configuration
(\ref{line_el_scha}) by integrating the energy density in the whole
spacetime volume where the Killing vector is well defined, that is,
from the event horizon $r_H$ up to the spatial infinity

\begin{equation}
\label{int_en_tot}
E_{g}=4 \pi \, \int^{\infty}_{r_{H}} \, \mathcal{E}_{g}(r) \, r^{2} \, dr,
\end{equation}
where $\mathcal{E}_{g}$ is solution of equation (\ref{220819}).
However, this integral cannot be done analytically and, therefore,
we appeal to numerical methods for the calculation of the total
energy. Using the open-source SciPy$^{\copyright}$ \cite{num}, we obtain $E_{g}\gtrapprox0.8\,Mc^2$ as an estimation for the integral\ (\ref{int_en_tot}).

\section{Conclusion}
The formula of $ t^{g}_{\mu\nu} $ proposed here is a consequence of
the description of the gravitational field in terms of the unique
scalar function $ \Phi.$ We used the field theoretical approach as
it was done in the case of General Relativity by Feynman et al., once
it seems that it is the most direct way to bypass the difficulties
invoked by many authors based on the equivalence principle. We have
shown that it is possible to define a true energy-momentum tensor
$t^{\mu\nu}_{g}$ whose conservation implies the dynamical
equations of the field. In this paper we only start the program to
examine the behavior of this quantity in stars. There remains the
task to analyze further the complete properties of $t^{g}_{\mu\nu}$
and to apply it in others configurations. This is a matter for
future work.

\appendix
\section{Particular choice of $a$ and $b$ from the variational principle}\label{app1}
Let us explain here the origin of the particular values of $ a $ and
$ b$ proposed in the text to specify the tensor $ \Theta_{\mu\nu}.$

Start by re-calling the action $$ I = \frac{1}{\kappa} \, \int \,
\sqrt{- \eta} \, V \, w $$ The dynamics of $ \Phi$ is obtained
directly by varying $ I. $ However this procedure can be realized in
two stages: varying the action with respect to the gravitational
metric $ q_{\mu\nu}$ it gives which components of the matter tensor
will be taken into account in the dynamics, obtaining Eq.\
(\ref{20081}), and then, varying $q_{\mu\nu}$ with respect to $
\Phi$ we get the dynamics (\ref {12out1}).

Let us use the definition of the gravitational metric $ q_{\mu\nu} $
and rewrite this action in the equivalent manner
\begin{equation}
I = \frac{1}{\kappa} \, \int \, \sqrt{- q} \, \sqrt{V} \,
\partial_{\mu} \Phi \,
\partial_{\nu} \Phi \, q^{\mu\nu} \label{22082}
\end{equation}
The dynamics of $ \Phi$ is obtained by variation of $ I$ with
respect to arbitrary variations $ \delta \Phi.$ In a first step we
vary with respect to $ \delta q^{\mu\nu}.$   We thus obtain the
intermediary variation
$$ \delta I = \frac{1}{\kappa} \, \int \sqrt{- q} \, \sqrt{V} \, \left( \partial_{\mu} \Phi
\, \partial_{\nu} \Phi - \frac{1}{2} \, \Omega \, q_{\mu\nu} \right)
\, \delta q^{\mu\nu}. $$

This suggests to define the associated $ \Theta_{\mu\nu} $ using
equation (\ref{22084}) up to a factor $2$:

$$ \Theta_{\mu\nu}= \frac{1}{\sqrt{-q}} \, \frac{\delta (\sqrt{-q}
\, \sqrt{V} \, \partial_{\mu} \Phi \,
\partial_{\nu} \Phi \, q^{\mu\nu})
}{\delta q^{\mu\nu}} \, .$$

It then follows
\begin{equation}
\Theta_{\mu\nu} = - \, 4 \, \sqrt{V} \, \left(
\partial_{\mu} \Phi \, \partial_{\nu} \Phi - \frac{1}{2} \, \Omega
\, q_{\mu\nu} \right). \label{22086}
\end{equation}

This expression yields the values of $ a $ and $ b$ that we used in
the original expression of $ \Theta_{\mu\nu}.$\\

\section{The energy density of a compact object}\label{app2}
We shall give the steps to solve the differential equation (\ref{220819}) for the energy density. Due to the term $|\alpha-3|$ in the function $P(\alpha)$, we separate $\mathcal{E}_{g}$ in two regimes: $\mathcal{E}_{g,1}$ for $1<\alpha<3$ and $\mathcal{E}_{g,2}$ for $\alpha>3$. The general solution of (\ref{220819}) is easily obtained if we do the change of variables $z=\sqrt[3]{\alpha-1}$. Therefore, for $0<z<\sqrt[3]{2}$, we have
\begin{widetext}
\begin{equation}
\label{en_d_z0}
\begin{array}{lcl}
\mathcal{E}_{g,1}(z)&=&-\fracc{1}{6Kr_H^2} \left[ -\fracc{9 z^2}{2(z^3+1)^4} + \fracc{54 z^2}{(z^3+1)^3} - \fracc{6 z^2}{(z^3+1)^2} - \fracc{z^2}{9(z^3+1)} + \fracc{7 \sqrt{3}}{27}\arctan\left(\fracc{-1 + 2 z}{\sqrt{3}}\right)+\right.\\[2ex]
&&\left.- \fracc{2^{8/3}\sqrt{3}}{27}\arctan\left(\fracc{1 + 2^{2/3} z}{\sqrt{3}}\right) - \fracc{7}{27} \ln(1 + z) - \fracc{2^{8/3}}{27}\ln(2 - 2^{2/3} z) + \fracc{7}{54}\ln(1 - z + z^2)+\right.\\[2ex]
&&\left.+\fracc{2^{8/3}}{27}\ln(2 + 2^{2/3} z + 2^{1/3} z^2)-\fracc{7 \sqrt{3}}{27}\arctan\left(-\fracc{\sqrt{3}}{3}\right)+\fracc{\sqrt{3}2^{8/3}}{27}\arctan\left(\fracc{\sqrt{3}}{3}\right)\right] z^4\, (z^3+1)^{-1/2}.
\end{array}
\end{equation}
\end{widetext}
where the constant of integration is chosen in such a way that $\mathcal{E}_{g,1}$ vanishes when $z$ goes to zero (which means $r\rightarrow\infty$). For $z>\sqrt[3]{2}$, we obtain that $\mathcal{E}_{g,2}=-\mathcal{E}_{g,1}$. Finally, the total energy in given by
\begin{widetext}
$$E = 12\pi \int_{0}^{\infty} \mathcal{E}_{g}(z) \fracc{(z^3+1)^2}{z^{10}}dz = 12\pi \left( \int_{0}^{\sqrt[3]{2}} \mathcal{E}_{g,1}(z) \fracc{(z^3+1)^2}{z^{10}}dz +\int_{\sqrt[3]{2}}^{\infty}\mathcal{E}_{g,2}(z)\fracc{(z^3+1)^2}{z^{10}}dz\right).$$
\end{widetext}
\section*{Acknowledgements} We would like to thank J.M. Salim and N. Pinto-Neto for their interest and important comments on this paper and G.B. Caminha who helped us with the numerical calculations. This work was supported by CNPq and FAPERJ.


\begin{thebibliography}{100}
\bibitem{2}
We note that distinct from GR in the Scalar Gravity only five components of the energy-momentum of matter enter in the dynamics of the gravitational field. We used this property to re-write the dynamics of GSG in terms of a field theory in the auxiliary Minkowski background along the same lines as it was undertaken by Gupta, Feynmann and others in the realm of GR.
\bibitem{num}
Some parts of the numerical integration were also verified making use of Maple$^{\copyright}$\, and Mathematica$^{\copyright}$.
\end{thebibliography}
\end{document}